\documentstyle[12pt,german]{article}
\begin{document}
\hfill LMU--06/97
\begin{center}
{\bf FLAVOR MIXING AND THE\\
MASSES OF THE LIGHT QUARKS\renewcommand{\thefootnote}{\fnsymbol{footnote}}
\footnote[1]{Talk given at the 1997 Workshop on Masses and
Mixings of Quarks and Leptons, Shizuoka, Japan (March 1997)}\renewcommand{\thefootnote}{\fnsymbol{footnote}} 
\footnote[2]{Supported by DFG--contract 412/22--1, EEC--contract SC1-CT91-0729, 
EEC--contract CHRX--CT94--0579 (DG 12 COMA)}}
\end{center}
\begin{center}
{\small HARALD FRITZSCH}\\
\vspace*{3mm}
{\small \it Yukawa Institute of Theoretical Physics, Kyoto, Japan}\\
\vspace*{3mm}
and\\
\vspace*{3mm}
{\small \it Ludwig--Maximilians--Universit"at, Sektion Physik,}\\
{\small Theresienstrasse 37, D--80333 M"unchen}\\
\bigskip
{\footnotesize E--mail: bm@hep.physik.uni-muenchen.de}
\end{center}
\vspace*{0.2cm}
\begin{center}
ABSTRACT\\
\vspace*{0.1cm}
\parbox[t]{12cm}{\footnotesize \hspace*{0.5cm}
A simple breaking of the subnuclear democracy of the quarks leads to a mixing
between the second and the third family, in agrement with observation.
Introducing the mixing between the first and the second family, one finds
an interesting pattern of maximal $CP$--violation as well as a complete
determination of the elements of the CKM matrix and of the unitarity
triangle.}
\end{center}
\bigskip
\hspace*{0.5cm}
In the standard electroweak model both the masses of the quarks as well
as
the weak mixing angles enter as free parameters, given by the associated 
Yukawa coupling constants. Any further insight
into
the yet unknown dynamics of mass generation would imply a step beyond
the
 physics
of the electroweak standard model. At present it seems far too early to
attempt an actual solution of the dynamics of mass generation, and one
is
invited to follow a strategy similar to the one which led eventually to
the
solution of the strong interaction dynamics by QCD, by looking for
specific patterns and symmetries as well as specific symmetry
violations in the internal flavor space of the quarks and leptons.\\
\\
\indent
The mass spectra of the quarks are dominated largely by the masses of
the
members of the third family, i.\ e.\ by $t$ and $b$. Thus a clear
hierarchical
pattern exists. Furthermore the masses of the first family are small
compared
to those of the second one. Moreover, the CKM--mixing matrix exhibits a
hierarchical pattern -- the transitions between the second and third
family
as well as between the first and the third family are small compared to
those between the first and the second family.\\
\newpage
\setcounter{page}{1}
\pagestyle{plain}
About 15 years ago, it was emphasized$^{1)}$ that the observed
hierarchies signify
that nature seems to be close to the so--called ``rank--one'' limit, in
which
all mixing angles vanish and both the u-- and d--type mass matrices are
proportional to the rank-one matrix
\begin{equation}
M_0 = {\rm const.} \cdot \left(\begin{array}{ccc}
0 & 0 & 0\\ 0 & 0 & 0\\ 0 & 0 & 1 \end{array} \right) \, .
\end{equation}
\indent
Whether the dynamics of the mass generation allows that this limit can
be
achieved in a consistent way remains an unsolved issue, depending on the 
dynamical details of mass generation. 
In this paper we take the point of view that the quark mass eigenvalues are
dynamical entities, and one could change their values in order to study
certain symmetry limits, as it is done in QCD. In the standard electroweak
model, in which the quark mass matrices are given by the coupling of a
scalar field to the various quark field, this can certainly be done by
changing the related--coupling constants. Whether it is possible in reality
remains to be seen.\\
\\
\indent
It is well--known that the quark mass matrices can always be made hermitean
by a suitable transformation of the righthanded fields. We shall suppose in
this paper that the quark mass matrices are hermitean. The limit described by
eq.\ (1) is a non--trivial constraint; it can be derived from imposing a
chiral symmetry, as emphasized in ref. (2).
This symmetry ensures that an electroweak doublet which is massless
remains
unmixed and is coupled to the $W$--boson with full strength.\\
\\
\indent
As soon as
the mass is introduced, at least for one member of the doublet, the symmetry
is
violated and mixing phenomena are expected to show up. That way a chiral
evolution of the CKM matrix can be constructed.$^{2)}$ At the first stage
only the $t$ and $b$ quark masses are introduced, due to their
non-vanishing
coupling to the scalar ``Higgs'' field. The CKM--matrix is unity in this
limit. At the next stage the second generation acquires a mass.
Since
the $(u, d)$--doublet is still massless, only the second and the third
 generations
mix, and the CKM--matrix is given by a real $2 \times 2$ rotation matrix
in the
$(c, \, s) - (t, \, b)$ subsystem, describing e.\ g.\ the mixing between
$s$
and $b$.\\
\\
\indent
In the limit where the masses of the $u$ and $d$ quarks are set to zero, the
quark mass matrices $M_{ij}$ both for the charge $2/3$ and charge $- 1/3 $ 
quarks can be arranged such that all elements  $M_{ij}$ and
$M_{ij} (i = 1, 2, 3)$ are zero.\\
\\
Thus the quark mass matrices have the form:\\
\begin{equation}
M_{ij} =
\left(
\begin{array}{clc}
0 & 0   & 0\\
0 & a   & b\\
0 & b^* & C
\end{array} \right) 
\end{equation}
The observed mass hierarchy is incorporated into this structure by denoting
the entry which is of the order of the $t$-- or $b$--mass by $C$, which
$a, \mid b \mid << C$. It can easily be seen (see, e.\ g.\ ref.\ (3) that the
complex phases in the matrices given in e.\ g.\ (1) can be rotated away by
subjecting both $M_{ij}^u$ and $M_{ij}^d$ to the same unitary transformation.
Thus we shall take $b$ to be real both for $U$--quarks and for $D$--quarks.
As expected, $CP$--violation cannot arise at this stage.\\
\\
\indent
Only at the next step, at which the $u$ and $d$ masses are
introduced, does the full CKM--matrix appear, described in general by
three angles
 and one
phase, and only at this step $CP$--violation can appear. Thus it is the
generation of mass for the first family which is responsible for the
violation of $CP$--symmetry.\\
\\
\indent
It has been emphasized some time ago$^{4, \, 5)}$ that the rank-one mass
matrix (see
 eq.
(1)) can be expressed in terms of a ``democratic mass matrix'':
\begin{equation}
M_0 = c \left( \begin{array}{ccc}
1 & 1 & 1\\ 1 & 1 & 1\\ 1 & 1 & 1 \end{array} \right) \, ,
\end{equation}
which exhibits an $S(3)_L \, \, \times \, \, S(3)_R$ symmetry. Writing
down
the mass eigenstates in terms of the eigenstates of the
``democratic'' symmetry, one finds e.g. for the $u $--quark channel:
\begin{eqnarray}
u^0 & = & \frac{1}{\sqrt{2}} (u_1 - u_2) \nonumber\\
c^0 & = & \frac{1}{\sqrt{6}} (u_1 + u_2 - 2u_3)\\
t^0 & = & \frac{1}{\sqrt{3}} (u_1 + u_2 + u_3) \nonumber .
\end{eqnarray}
Here $u_1, \ldots$ are the symmetry eigenstates.
Note that $u^0$ and $c^0$ are massless
in
the limit considered here, and any linear combination of the first two
state vectors given in eq. (3) would fulfill the same purpose, i.\ e.\
the
decomposition is not unique, only the wave function of the coherent
state
$t^0$ is uniquely defined. This ambiguity will disappear as soon as
the symmetry is violated.\\
\\
\indent
The wave functions given in eq. (3) are reminiscent of the wave
functions
of the neutral pseudoscalar mesons in QCD in the $SU(3)_L \, \, \times
\, \,
SU(3)_R$ limit:
\begin{eqnarray}
\pi^0_0 & = & \frac{1}{\sqrt{2}}(\bar uu - \bar dd)\\
\eta_0 & = & \frac{1}{\sqrt{6}}(\bar uu + \bar dd - 2\bar ss)
\nonumber\\
\eta '_0 & = & \frac{1}{\sqrt{3}}(\bar uu + \bar dd + \bar ss) .
\nonumber
\end{eqnarray}
(Here the lower index denotes that we are considering the chiral limit).
Also the mass spectrum of these mesons is identical to the mass spectrum
of
the quarks in the ``democratic'' limit: two mesons $(\pi
^0_0 \, ,
\eta _0)$ are massless and act as Nambu--Goldstone bosons, while the
third
coherent state $\eta '_0$ is \underline{not} massless due to the QCD
anomaly.\\
\\
\indent
In the chiral limit the (mass)$^2$--matrix of the neutral pseudoscalar
mesons
is also a ``democratic'' mass matrix when written in terms of the $(\bar
qq)$--
eigenstates $(\bar uu), \, (\bar dd)$ and $(\bar ss) \, ^{6)}$:
\begin{equation}
M^2(ps) = \lambda \left( \begin{array}{ccc}
1 & 1 & 1\\ 1 & 1 & 1\\ 1 & 1 & 1 \end{array} \right),
\end{equation}
where the strength parameter $\lambda $ is given by
$\lambda = M^{2}(\eta '_{0}) \, / \, 3$.
The mass matrix (6) describes the result of the QCD--anomaly which
causes strong
transitions between the quark eigenstates (due to gluonic annihilation
effects
enhanced by topological effects). Likewise one may argue that analogous
transitions are the reason for
the lepton--quark mass hierarchy. Here we shall not speculate about a
detailed mechanism of this type, but merely study the effect of symmetry
breaking.\\
\\
\indent
In the case of the pseudoscalar mesons the breaking of
the symmetry down to
$SU(2)_L \, \, \times \, \, SU(2)_R$ is provided by a direct mass term
$m_s \bar
 ss$
for the s--quark. This implies a modifica\-tion of the (3,3) matrix
element in
eq. (5), where $\lambda $ is replaced by $\lambda + M^2(\bar ss)$ where
 $M^2(\bar ss)$ is
given by $2M^2_K$, which is proportional to $< \bar ss >_0$, the
expectation
value of $\bar ss$ in the QCD vacuum. This direct mass term causes the
violation of the symmetry and generates at the same time a mixing
between
$\eta _0$ and $\eta '_{0}$, a mass for the $\eta _{0}$, and a mass shift
for
the $\eta '_{0}$.\\
\\
\indent
It would be interesting to see whether an analogue of the
simplest violation of this kind of symmetry violation of the ``democratic'' 
symmetry which describes successfully
the mass and mixing pattern of the $\eta - \eta '$--system is also able to
describe the observed mixing and mass pattern of the second and third
family of leptons and quarks. This was discussed recently$^{7)}$.
Let us replace the (3,3) matrix element in
eq.\ (2) by $1 + \varepsilon _i$; (i = u (u--quarks), d
(d--quarks)
respectively. The small real parameters $\varepsilon _i$ describe the
departure
from democratic symmetry and lead
\begin{enumerate}
\item[a)]
to a generation of mass for the second family and
\item[b)] to a flavour mixing between the third and the second
family. Since $\varepsilon $ is directly related (see below) to a
fermion mass
and the latter is \underline{not} restricted to be positive,
$\varepsilon $
can be positive or negative. (Note that a negative Fermi--Dirac mass can
always be turned into a positive one by a suitable $\gamma
_5$--transformation
of the spin $\frac{1}{2}$ field). Since the original mass term is
represented by a symmetric matrix, we take $\varepsilon $ to be real.
\end{enumerate}
\hspace*{0.5cm}
In ref.\ [5] a general breaking of the flavor democracy was discussed in
term of two parameters $\alpha $ and $\beta $. The ansatz discussed here,
in analogy to the case of the pseudoscalar mesons which represents the
simplest breaking of the flavor democracy, corresponds to the special case
$\alpha = 0$. Note that the case $\beta = \alpha + \alpha^*$
discussed in ref.\ [4] leads to the mass matrix given in ref.\ [1].\\
\\
\indent
It is instructive to rewrite the mass matrix in the hierarchical basis,
where
one obtains in the case of the down--type quarks:\\
\begin{equation}
M = c_l \left( \begin{array}{ccc}
0 & 0 & 0\\ 0 & + \frac{2}{3}{\varepsilon _u} & - \frac{\sqrt{2}}{3}
\varepsilon_u\\
0 & - \frac{\sqrt{2}}{3} \varepsilon _u & 3 + \frac{1}{3}\varepsilon _u
 \end{array}
\right) \, .
\end{equation}
\indent
In lowest order of $\varepsilon $ one finds the mass eigenvalues
$m_s = \frac{2}{9} \varepsilon _d \cdot m_b \, , m_b = m_{b^0} \, ,
\Theta _{s, b} = | \sqrt{2} \cdot \varepsilon _d / 9|$.\\
\\
\indent
The exact mass eigenvalues and the mixing angle are given by:
\begin{eqnarray}
m_1 / c_d & = & \frac{3 + \varepsilon _d}{2} - \frac{3}{2} \sqrt{1 -
\frac{2}{9} \varepsilon _d + \frac{1}{9} \varepsilon _d ^2} \nonumber\\
m_2 / c_d & = & \frac{3 + \varepsilon _d}{2} + \frac{3}{2} \sqrt{1 -
\frac{2}{9}
\varepsilon _d + \frac{1}{9} \varepsilon _d^2}\\
\sin \Theta _{(s,b)} & = & \frac{1}{\sqrt{2}}\left( 1 - \frac{1 -
 \frac{1}{9}\varepsilon _d}
{(1 - \frac{2}{9} \varepsilon _d + \frac{1}{9} \varepsilon
 ^2_d)^{1/2}}\right)^{1/2} \nonumber \, .
\end{eqnarray}
\indent
The ratio $m_s / m_b$ is allowed to vary in the range
$0.022 \ldots 0.044$ (see ref. (8)). According to eq. (7) one finds
$\varepsilon _d$ to vary from $\varepsilon _d = 0.11$ to $0.21$.
The associated $s - b$
mixing angle varies from $\Theta (s, b) = 1.0^{\circ }$ \hspace{0.3cm}
$(\sin \Theta = 0.018)$ and $\Theta (s, b) = 1.95^{\circ }$
\hspace{0.3cm}
$(\sin \Theta = 0.034)$. As an illustrative example we use the values
$m_b(1GeV)
 = 5200 MeV$, \hspace{0.3cm}
$m_s(1GeV) = 220 MeV$. One obtains $\varepsilon _d = 0.20$ and
$\sin \Theta(s, b) = 0.032$.\\
\\
\indent
To determine the amount of mixing in the $(c, t)$--channel, a knowledge
of
the ratio $m_c / m_t$ is required. As an illustrative example we take
$m_c(1GeV) = 1.35 GeV$, $m_t(1GeV) = 260 GeV$ (i.\ e.\ $m_t(m_t) \approx
160GeV)$,
which gives $m_c / m_t \cong 0.005$. In this case one finds $\varepsilon
_u =
 0.023$ and $\Theta(c,t) = 0.21^{\circ }$
\hspace{0.3cm} $(\sin \Theta (c, t) = 0.004$) .\\
\\
\indent
The actual weak mixing between the third and the second quark family is
combined effect of the two family mixings described above. The symmetry
breaking given by the $\varepsilon $--parameter can be interpreted, as
done
in eq. (7), as a direct mass term for the $u_3, d_3$ fermion.
However, a direct fermion mass term need not be positive, since its sign
can always be changed by a suitable $\gamma _5$--transformation. What
counts
for our ana\-lysis is the relative sign of the $m_s$--mass term in
comparison
to the $m_c$--term, discussed previously. Thus two possibilities must be
considered:
\begin{enumerate}
\item[a)] Both the $m_s$-- and the $m_c$--term
have the same relative sign with respect to each other, i.\ e.\ both
 $\varepsilon _d$
and $\varepsilon _u$ are positive, and the mixing angle between the
second
and third family is given by the difference $\Theta (sb) - \Theta (ct)$.
This
possibility seems to be ruled out by experiment, since it would lead to
$V_{cb} < 0.03$.
\item[b)] The relative signs of the breaking
terms
$\varepsilon _d$ and $\varepsilon _u$ are different, and the mixing
angle
between the $(s,b)$ and $(c,t)$ systems is given by the sum
$\Theta(sb) + \Theta(ct)$. Thus we obtain $V_{cb} \cong \sin (\Theta(sb)
+ \Theta(ct))$.
\end{enumerate}
\hspace*{0.5cm}
According to the range of values for $m_s$ discussed above, one finds
$V_{cb} \cong 0.022 ... 0.038$.
For example, for $m_s(1GeV) = 220MeV$, \, $m_c (1GeV) = 1.35 GeV$,
$m_t(1GeV)
= 260GeV$ one obtains $V_{cb} \cong 0.036$.\\
\\
\indent
The experiments give $V_{cb} = 0.032 \dots 0.048^{9)}$. We conclude from
the analysis
 given above
that our ansatz for the symmetry breaking reproduces the lower part of
the
experimental range. Nevertheless we obtain consistency with experiment only
if the ratio $m_s / m_b$ is relatively large implying $m_s(1GeV) \ge
180MeV$. Note that recent estimates of $m_s$ (1GeV) give values in the
range 180 $ \ldots $ 200 MeV$^{10)}$.\\
\\
\indent
It is remarkable that the simplest ansatz for the breaking of the
``democratic
symmetry'', one which nature follows in the case of the pseudoscalar
mesons, is able to reproduce the experimental data on the mixing between
the second and third family. We interpret this as a hint that the
eigenstates
of the symmetry, not the mass eigenstates,
play
a special r\^{o}le in the physics of flavour, a r\^{o}le
which needs to be investigated further.\\
\\
\indent
The next step is to introduce the mass of the $d$ quark, but
keeping $m_{u}$ massless. We regard this sequence of steps
as useful due to the
fact that the mass ratios $m_{u}/m_{c}$ and $m_{u}/m_{t}$ are about one order
of magnitude smaller than the ratios $m_{d}/m_{s}$ and $m_{d}/m_{b}$
respectively.
It is well-known that the observed magnitude of the mixing between the
first and the second family can be reproduced well by a specific texture
of the mass matrix [11, 12]. We shall incorporate this here and take the following
ansatz for the mass matrix of the down-type quarks:\\
\begin{equation}
M_{\rm d} \; = \; \left ( 
\begin{array}{ccc}
0  & D_{\rm d}  & 0 \\
D^{*}_{\rm d}  & C_{\rm d}  & B_{\rm d} \\
0  & B_{\rm d}  & A_{\rm d}
\end{array}
\right ) \; .
\end{equation}
The consideration above with respect to the breaking of the democratic
symmetry suggest that $C_d / B_d = - \sqrt{2}$. However, for our subsequent
consideration this specific ratio is not essential.
At this stage the mass matrix of the up-type quarks remains in the form (6).
The CKM matrix elements $V_{us}$, $V_{cd}$ and the ratios $V_{ub}/V_{cb}$, $V_{td}/V_{ts}$
can be calculated in this limit. One finds in lowest order:
\begin{equation}
V_{us} \; \approx \; \sqrt{\frac{m_{d}}{m_{s}}} \; , \;\;\;\;\;\;\;
V_{cd} \; \approx \; \sqrt{\frac{m_{d}}{m_{s}}} \; , \;\;\;\;\;\;\;
\frac{V_{ub}}{V_{cb}} \; \approx \; 0 \; , \;\;\;\;\;\; \frac{V_{td}}{V_{ts}} \; \approx
\; \sqrt{\frac{m_{d}}{m_{s}}} \; .
\end{equation}
\indent
An interesting implication of the ansatz (8) is the vanishing of $CP$ violation.
Although the mass matrix (5) contains a complex parameter $D_d$, its phase
can be
rotated away due to the fact that $m_{u}$ is still massless, and a phase rotation
of the $u$-field does not lead to any observable consequences. The vanishing
of $CP$ violation can be seen as follows. Considering two hermitian mass matrices
$M_{\rm u}$ and $M_{\rm d}$ in general, one may define a commutator like
\begin{equation}
\left [ M_{\rm u}, M_{\rm d} \right ] \; = \; i{\cal C} \; 
\end{equation}
and prove that its determinant ${\rm Det}~{\cal C}$ is a rephasing invariant 
measure of $CP$ violation [13].
It can easily be checked that Det C vanishes.
The vanishing of $CP$ violation in our approach in the limit
$m_{u}\rightarrow 0$
is an interesting phenomenon, since it is the same limit in which the
``strong''
$CP$ violation induced by instanton effects of QCD is absent [14]. Whether
this link
between ``strong'' and ``weak'' $CP$ violation could offer a solution of the 
``strong'' $CP$ problem remains an open issue at the moment. Nevertheless it
is
an interesting feature of our approach that $CP$ violation and the mass of
the $u$ quark
are intrinsically linked to each other. Since the phase of $D$ can be rotated
away, it will be disregarded, and $D$ is taken to be real.\\
\\
\indent
The final step is to introduce the mass of the $u$ quark. The mass matrix 
$M_{\rm u}$ takes the form:
\begin{equation}
M_{\rm u} \; = \; \left ( 
\begin{array}{ccc}
0  & D_{\rm u}  & 0 \\
D^{*}_{\rm u}  & C_{\rm u}  & B_{\rm u} \\
0  & B_{\rm u}  & A_{\rm u}
\end{array}
\right ) \; .
\end{equation}
(Here $A_u$ etc.\ are defined analogously as in e.g.\
(8)).
Once the mixing term $D_{\rm u}=|D_{\rm u}|e^{i\sigma}$ for the $u$-quark is introduced, $CP$ violation
appears. For the determinant of the commutator (6) we find$^{15)}$:
\begin{equation}
{\rm Det}~{\cal C} \; \cong \; T \sin \sigma ,
\end{equation}
\begin{equation}
\begin{array}{lll}
T & = & 2 |D_{\rm u}D_{\rm d}| \left [ (A_{\rm u}B_{\rm d}-B_{\rm u}A_{\rm d})^{2}
-|D_{\rm u}|^{2}B_{\rm d}^{2}-B_{\rm u}^{2}|D_{\rm d}|^{2} \right . \\
&  &    \left . -(A_{\rm u}B_{\rm d}-B_{\rm u}A_{\rm d})(C_{\rm u}B_{\rm d}-B_{\rm u}C_{\rm d})\right ] \; .\\
\end{array}
\end{equation}
\indent 
The phase $\sigma$ determines the strength of $CP$ violation.
The diagonalization of the mass matrices $M_{\rm d}$ and $M_{\rm u}$ leads to theigenvalues $m_{i}$ ($i=u,d,...$). Note that $m_{u}$ and $m_{d}$ appear to be negative.
By a suitable $\gamma_{5}$-transformation of the quark fields one can arrange them to be
positive. Collecting the lowest order terms in the CKM matrix, one obtains:
\begin{equation}
V_{us} \; \approx \; \sqrt{\frac{m_{d}}{m_{s}}} -\sqrt{\frac{m_{u}}{m_{c}}}e^{i\sigma} \; , \;\;\;\;\;\;\;
V_{cd} \; \approx \; \sqrt{\frac{m_{u}}{m_{c}}} -\sqrt{\frac{m_{d}}{m_{s}}}e^{i\sigma} \; 
\end{equation}
and
\begin{equation}
\frac{V_{ub}}{V_{cb}} \; \approx \; - \sqrt{\frac{m_{u}}{m_{c}}} \; , 
\;\;\;\;\;\;\;\;\; \frac{V_{td}}{V_{ts}} \; \approx \; - \sqrt{\frac{m_{d}}{m_{s}}} \; .
\end{equation}

\indent
The relations for $V_{us}$ and $V_{cd}$ were obtained previously [12].
However then it was not noted that the relative phase between the two ratios might be 
relevant for $CP$ violation. A related discussion can be found in ref.\ [16].\\
\\
\indent
According to eq.\ (12) the strength of $CP$ violation depends on the phase 
$\sigma$. If we keep the modulus of the parameter
$D_{\rm u}$ constant, but vary the phase from zero to
$90^{0}$, the
strength of $CP$ violation varies from zero to a maximal value given by eq.
(12), which is obtained for $\sigma = 90^{\circ}$.
We conclude that 
$CP$ violation is maximal for $\sigma =90^{0}$.
In this case the element
$D_{\rm u}$ would be purely imaginary, if we set the phase of the matrix element
$D_{\rm d}$ to be zero. As discussed above, this can always be arranged.\\
\\
\indent
In our approach the $CP$-violating phase also enters in the expressions for $V_{us}$
and $V_{cd}$ (Cabibbo angle). As discussed already in ref.\ [12], the Cabibbo angle
is fixed by the difference of $\sqrt{m_{d}/m_{s}}$ and $\sqrt{m_{u}/m_{c}}$ $\times$ phase factor.
The second term contributes a small correction (of order 0.06) to the leading
term, which according to the mass ratios given in ref. [8] is allowed to vary between
0.20 and 0.24. For our subsequent discussion we shall use
$0.218\leq |V_{us}|\leq 0.224$ [8].
If the phase parameter multiplying $\sqrt{m_{u}/m_{c}}$ were zero or $\pm 180^{0}$
(i.e. either the difference or sum of the two real terms would enter), the observed magnitude
of the Cabibbo angle could not be reproduced. Thus a phase is needed, and we find within
our approach purely on phenomenological grounds that $CP$ violation must be present if
we request consistency between observation and our result (14).\\
\\
\indent
An excellent description of the magnitude of $V_{us}$ is obtained for a phase angle of $90^{0}$.
In this case one finds:
\begin{equation}
|V_{us}|^{2} \; \approx \; \left (1-\frac{m_{d}}{m_{s}}\right )
\left (\frac{m_{d}}{m_{s}}+\frac{m_{u}}{m_{c}}\right ) \; ,
\end{equation}
where approximations are made for $V_{us}$ to a better degree of accuracy
than that in eq. (14).
Using $|V_{us}|$ = 0.218...0.224 and $m_{u}/m_{c}$ = 0.0028...0.0048 we obtain
$m_{d}/m_{s}$ $\approx$ 0.045...0.05. This corresponds to $m_{s}/m_{d}$ $\approx$ 20...22,
which is entirely consistent with the determination of $m_{s}/m_{d}$, based on chiral
perturbation theory [8]: $m_{s}/m_{d}$ = 17...25. This example shows that the phase angle 
must be in the vicinity of $90^{0}$.
Fixing $m_{u}/m_{c}$ to its central value and varying $m_{d}/m_{s}$ throughout the 
allowed range, we find $\sigma \approx 66^{0}...110^{0}$.\\ 
\\
\indent
The case $\sigma=90^{0}$, favoured by our analysis, deserves a special
attention. It implies that in the sequence of steps discussed above the term
$D_{\rm u}$ generating the mass of the $u$-quark is purely imaginary, and hence
$CP$ violation is maximal. It is of high interest to observe that nature seems to prefer
this case. A purely imaginary term $D_{\rm u}$ implies that the algebraic structure
of the quark mass matrix is particularly simple. Its consequences need to be 
investigated further and might lead the way to an underlying internal symmetry
responsible for the pattern of masses.\\
\\
\indent
Finally we explore the consequences of our approach to the unitarity triangle,
i.e., the triangle formed by the CKM matrix elements $V^{*}_{ub}$, $V_{td}$
and $s_{12}V_{cb}$ ($s_{12}=\sin\theta_{12}$, $\theta_{12}$: Cabibbo angle)
in the complex plane (we shall use the definitions of the angles 
$\alpha$, $\beta$ and $\gamma$ as given in ref. [9). For $\sigma=90^{0}$ we
obtain:\\
\begin{equation}
\alpha \approx 90^{\circ }, \qquad \beta \approx {\rm arctan} \sqrt{\frac{m_u}{m_c}
\cdot \frac{m_s}{m_d}} \, , \qquad \gamma \approx 90^{\circ } - \beta \, .
\end{equation}
\indent
Thus the unitarity triangle is a rectangular triangle. 
We note that
the unitarity triangle and the triangle formed in the complex phase by $V_{us}$,
$\sqrt{m_d / m_s}$ and $\sqrt{m_u / m_c}$ are similar rectangular triangles,
related by a scale transformation. Using as input
$m_{u}/m_{c}$ = 0.0028...0.0048 and $m_{s}/m_{d}$ = 20...22 as discussed above,
we find $\beta$ $\approx$ $13^{0}...18^{0}$, $\gamma$ $\approx$ $72^{0}...76^{0}$, 
and $\sin 2\beta$ $\approx$ $\sin 2\gamma$ $\approx$ 0.45...0.59.
These values are consistent with the experimental constraints.\\
\\
\indent
We have shown that a simple pattern for the generation of masses
for the first family of leptons and quarks leads to an interesting and
predictive pattern for the violation of $CP$ symmetry. The observed magnitude of the Cabibbo
angle requires $CP$ violation to be maximal or at least near to its maximal
strength.
The ratio $V_{ub}/V_{cb}$ as well as $V_{td}/V_{ts}$ are given by $\sqrt{m_{u}/m_{c}}$
and $\sqrt{m_{d}/m_{s}}$ respectively. In the case of maximal $CP$ violation the
unitarity triangle is rectangular ($\alpha=90^{0}$), the angle $\beta$ can vary in the
range $13^{0}...18^{0}$ ($\sin 2\beta=\sin 2\gamma$ $\approx$ 0.45...0.59). It remains
to be seen whether the future experiments, e.g. the measurements of the $CP$ asymmetry
in $B$--decays, $B^{0}_{d} $, confirm these values. \\
\newpage
\noindent
Acknowledgements:\\
\\
\indent
I would like to thank the staff of the Yukawa institute, especially Profs.\
T.\ Maskawa and M.\ Ninomiya, for the kind
hospitality extended to me during my stay in spring 1997.
\newpage
{\bf References}\\
\begin{enumerate}
\item H.\ Fritzsch, {\it Nucl.\ Phys.} {\bf B155} (1979) 189;\\
See also: H.\ Fritzsch, in: Proc.\ Europhysics Conf.\ on Flavor Mixing,\\
Erice, Italy (1984), Ling--Li Chau ed.

\item H.\ Fritzsch, {\it Phys.\ Lett.} {\bf B184} (1987) 391.

\item H.\ Lehmann and T.\ T.\ Wu, {\it Phys.\ Lett.} {\bf 384B} (1996) 249. 

\item H.\ Harari, Haut and J.\ Weyers,
{\it Phys.\ Lett.} {\bf 78B} (1978) 459;\\
Y.\ Chikashige, G.\ Gelmini, R.\ P.\ Peccei and M.\ Roncadelli,\\
{\it Phys.\ Lett.} {\bf 94B} (1980) 499;\\
C.\ Jarlskog, in: Proc.\ of the Int.\ Symp.\ on Production and Decay of\\
Heavy Flavors, Heidelberg, Germany, 1986;\\
P.\ Kaus and S.\ Meshkov, Mod.\ {\it Phys.\ Lett.} {\bf A3} (1988) 1251;
{\bf A4} (1989) 603;\\
G.\ C.\ Branco, J.\ I.\ Silva--Marcos, \, M.\ N.\ Rebelo,
{\it Phys.\ Lett.} {\bf B237} (1990) 446.
Y.\ Koide, {\it Phys.\ Rev.} {\bf D28} (1983) 252.

\item H.\ Fritzsch and J.\ Plankl,
{\it Phys.\ Lett.} {\bf B237} (1990) 451.

\item H.\ Fritzsch and P.\ Minkowski, {\it Nuovo Cimento}
{\bf 30A} (1975) 393;\\
H.\ Fritzsch and D.\ Jackson, {\it Phys.\ Lett.} {\bf 66B} (1977) 365.

\item H.\ Fritzsch and D.\ Holtmannsp\"otter,
{\it Phys.\ Lett.} {\bf B338} (1994) 290.

\item J.\ Gasser and H.\ Leutwyler, {\it Phys.\ Rev.} {\bf 87}
(1982) 77;\\
For a recent review of quark mass values, see, e.g., Y.\ Koide,
preprint US--94--05 (1994), and these Proceedings.

\item Particle Data Group, M.\ Aguilar--Benitez et al.,
{\it Phys.\ Rev.} {\bf D50} (1994) 1173.

\item M.\ Jamin and M.\ M\"unz, {\it Z.\ Phys.} {\bf C66} (1995) 633.

\item S.\ Weinberg, Transactions of the New York Academy of
Sciences,\\
Series II, Vol.\ 38, (1977) 185.

\item H.\ Fritzsch, {\it Phys.\ Lett.} {\bf 70B} (1977) 436.

\item C.\ Jarlskog, {\it Phys.\ Ref.\ Lett.} {\bf 55} (1984) 1039. 

\item R.\ D.\ Peccei and H.\ R.\ Quinn, {\it Phys.\ Rev.} {\bf D16}
(1977) 1791;\\
S.\ Weinberg, {\it Phys.\ Rev.\ Lett.} {\bf 40} (1978) 223;\\
F.\ Wilczek, {\it Phys.\ Rev.\ Lett.} {\bf 40} (1978) 279.

\item H.\ Fritzsch and Z.\ Xing, {\it Phys.\ Lett.} {\bf B353} (1995) 95.

\item M.\ Shin, {\it Phys.\ Lett.} {\bf B145} (1984) 285;\\
M.\ Gronau, R.\ Johnson and J.\ Schechter, {\it Phys.\ Rev.\ Lett.}
{\bf 54} (1985) 2176.
\end{enumerate}
\bigskip
\end{document}